\documentclass[12pt]{article}
\pdfoutput=1
\usepackage{subfigure}
\usepackage{amssymb,amsmath}
\usepackage{graphicx}
\usepackage{color}
\usepackage[colorlinks=true
,urlcolor=blue
,citecolor=blue
,linkcolor=blue
,pagecolor=blue
,linktocpage=true
,pdfproducer=medialab
]{hyperref}
\usepackage[a4paper,width=17cm]{geometry}
\makeatletter \renewcommand{\@dotsep}{10000} \makeatother
\usepackage{appendix}

\begin{document}

\begin{center}

 {\Large\bf  Non-Relativistic Limit of the Dirac Equation
 } \vspace{1cm}

{   Muhammad Adeel Ajaib\footnote{ E-mail: adeel@udel.edu}}

{\baselineskip 20pt \it
Department of Physics, California Polytechnic State University, San Luis Obispo, CA, 93401\\
 } \vspace{.5cm}

{\baselineskip 20pt \it
   } \vspace{.5cm}

\setcounter{footnote}{0}
\vspace{1.5cm}
\end{center}

\begin{abstract}

We show that the first order form of the Schr{\"o}dinger equation proposed in \cite{Ajaib:2015uha} can be obtained from the Dirac equation in the non-relativistic limit. We also show that the Pauli Hamiltonian is obtained from this equation by requiring local gauge invariance. In addition, we study the problem of a spin up particle incident on a finite potential barrier and show that the known quantum mechanical results are obtained. Finally, we consider the symmetric potential well and show that the quantum mechanical expression for the quantized energy levels of a particle is obtained with periodic boundary conditions. Based on these conclusions, we propose that the equation introduced in \cite{Ajaib:2015uha} is the non-relativistic limit of the Dirac equation and more appropriately describes spin 1/2 particles in the non-relativistic limit.

\end{abstract}

\newpage

\section{Introduction}\label{intro}

Schr{\"o}dinger equation lies at the foundations of our understanding of non-relativistic quantum mechanics. Although various scattering problems can be studied using this equation, it does not take into account the spin of the particle in these problems. The fundamental form of the Schr{\"o}dinger equation proposed in \cite{Ajaib:2015uha} allows for the inclusion of spin in scattering problems. It was shown in \cite{Ajaib:2015uha} that the proposed equation can be used to solve the finite step potential problem with the spin of the particle taken into account. The analysis therein predicts how a spin up electron scatters off a step potential. The transmission and reflection coefficient of the spin up and spin down particles were obtained and it was shown that, when added together, the coefficients of the spin up and down particle yield exactly the quantum mechanical result. In addition, a three dimensional version of the first order equation was also proposed in \cite{Ajaib:2015uha}.

In this article we derive results that further illustrates interesting consequences of this equation. We show that the first order form of the Schr{\"o}dinger equation can be obtained in the non-relativistic limit of the Dirac equation and the Pauli equation can be derived by requiring this equation to be locally invariant. We also show that the finite potential barrier problem can  be solved using the equation proposed in \cite{Ajaib:2015uha}. The analysis we perform takes into account the spin of the particle scattered from a finite potential barrier. In this case as well, we show that the sum of the transmission and reflection coefficients for the spin up and down particles yield the quantum mechanical result. Furthermore, we discuss the symmetric potential well problem also and show that energy quantization results from assuming periodic boundary conditions for this problem.  

The paper is organized as follows: In section \ref{nr-limit}, we show that the equation proposed in \cite{Ajaib:2015uha} is the non-relativistic equation of the Dirac equation. In section \ref{pauli-eq}, we derive the Pauli equation by requiring the first order Schr{\"o}dinger equation to be locally invariant. In section \ref{sec:fpb}, we analyze the finite potential barrier problem and section \ref{sec:box} discusses the problem of a particle in a symmetric potential well.  We conclude in section \ref{conclude}.

\section{The Non-Relativistic Limit of Dirac Equation}\label{nr-limit}

It was proposed in \cite{Ajaib:2015uha} that the Schr{\"o}dinger equation can be derived from a fundamental first order equation similar to the manner in which the Klein Gordon equation can be derived from the Dirac equation. 
The 3 dimensional version of the equation proposed in \cite{Ajaib:2015uha} is given by 
\begin{eqnarray}
-i \gamma_i \partial_i \psi = (i  \eta \partial_t  + \eta^\dagger m) \psi 
\label{mse-eq-3d}
\end{eqnarray}
where $\gamma_i$ are the Dirac matrices, $\eta$ is a 4$\times$4 nilpotent matrix and we choose $\hbar=c=1$. Here we use the representation $\eta=(\gamma_0+i \gamma_5)/\sqrt{2}$. There are several representations of the $\eta$  matrices and each  corresponds to a different representation of the gamma matrices. The representation $\eta=(\gamma_0+i \gamma_5)/\sqrt{2}$ corresponds to the standard representation of the Dirac gamma matrices.
We consider the following form of the Dirac equation\footnote{Equation (\ref{DE}) is equivalent to the standard Dirac equation. We can obtain the standard form of the Dirac equation by a simple redefinition of the field $\psi$ =  $ M\psi'$,
where 
$M = (1-i\gamma_5)/\sqrt{2}$
and then multiplying the equation with $M$ from the left.
}
\begin{eqnarray}
(i   \gamma^\mu \partial_\mu  - i \gamma_5 m) \psi =0
\label{DE}
\end{eqnarray}
or
\begin{eqnarray}
(i   \gamma_0 \partial_0+i \gamma_i \partial_i   - i \gamma_5 m) \psi =0
\end{eqnarray}
\vspace{2mm}
In momentum space ($\psi=u(p) e^{-i p.x}=u(p) e^{-i (E t - p_z z )}$) the above equation is given by
\begin{eqnarray}
( \gamma_0 E - \gamma_i p_i - i \gamma_5 m) u =0
\end{eqnarray}
using $\gamma_0=(\eta+\eta^\dagger)/\sqrt{2}$ and $i\gamma_5=(\eta-\eta^\dagger)/\sqrt{2}$, we get
\begin{eqnarray}
\left( \frac{\eta+\eta^\dagger}{\sqrt{2}} E - \gamma_i p_i - \frac{\eta-\eta^\dagger}{\sqrt{2}} m \right) u =0
\end{eqnarray}
\begin{eqnarray}
\left( \frac{\eta}{\sqrt{2}} (E-m) - \gamma_i p_i + \frac{\eta^\dagger}{\sqrt{2}} (E+m) \right) u =0
\end{eqnarray}
In the non-relativistic limit $E-m \simeq E'$ and $E+m \simeq 2m$, which yields
\begin{eqnarray}
\left( \frac{\eta}{\sqrt{2}} E' - \gamma_i p_i + \sqrt{2} \eta^\dagger m \right) u =0
\end{eqnarray}
which yields the dispersion relation of a non-relativistic particle ($E'=p_i^2/2m$). Equation (\ref{mse-eq-3d}) can be written in a more general form as
\begin{eqnarray}
\left( \frac{1}{a} \eta E - \gamma_i p_i + a \eta^\dagger m \right) u =0
\end{eqnarray}
or
\begin{eqnarray}
\left( i \eta^\mu \partial_\mu + a \eta^\dagger m \right) \psi =0 
\label{aeq}
\end{eqnarray}
where $a$ is a non-zero constant and $\eta^\mu=(\eta^0,\eta^i)=(\eta/a,\gamma^i)$. The above equation more appropriately describes spin 1/2 particles in the non-relativistic limit. An implication of this is that the above equation allows for inclusion of a particle's spin in the analysis of various problems in non-relativistic quantum mechanics.
 It was shown in reference \cite{Ajaib:2015uha} and we also show in section \ref{sec:fpb} that this equation allows for the inclusion of the spin of a particle in scattering problems. We will also show in section \ref{sec:box} that the symmetric potential well problem yields energy quantization from assuming periodic boundary conditions.

\section{Pauli Equation} \label{pauli-eq}
In the non-relativistic limit, spin can be introduced using the Pauli equation which describes the interaction of a spin 1/2 particle with an external electromagnetic field. It is obtained in the non-relativistic limit of the Dirac equation by assuming the presence of an electromagnetic field. It correctly predicts the spin of the particle and the gyromagnetic ratio. In this section we show that Pauli equation can obtained from equation (\ref{aeq}) by requiring local gauge invariance. We therefore require that the following equation\footnote{We choose $a=1$ for discussion in this section.}
\begin{eqnarray}
\psi^\dagger \left( i \eta^\mu \partial_\mu + \eta^\dagger m \right) \psi =0
\end{eqnarray}
is invariant under the local transformation
\begin{eqnarray}
\psi \rightarrow e^{-i e \theta(x)} \psi
\end{eqnarray}
This leads to the following locally invariant equation
%
%
%
%or
%
\begin{eqnarray}
\psi^\dagger \left( i \eta^\mu D_\mu + \eta^\dagger m \right) \psi =0
\end{eqnarray}
where $D_\mu=\partial_\mu+ i e A_\mu$, and the gauge field $A_\mu$ transforms as $A_\mu \rightarrow A_\mu+\partial_\mu \theta$ under the local transformation. In momentum space, the above equation is given by
\begin{eqnarray}
\psi^\dagger \left( \eta^\mu \Pi_\mu + \eta^\dagger m \right) \psi =0
\end{eqnarray}
where $\Pi_\mu= p_\mu -e  A_\mu$. Therefore, requiring gauge invariance amounts to performing the  transformation, $p_\mu \rightarrow p_\mu-e A_\mu$.
We obtain the Pauli Hamiltonian by squaring the above equation:
\begin{eqnarray}
H  &=&  \frac{(\vec \sigma.\vec \Pi)^2}{2m} +e A^0  \\
&=& \frac{\vec \Pi^2}{2m}- \frac{e}{2m} \vec \sigma.\vec B  +e A^0 
\label{dirac5}
\end{eqnarray}
where, $e$ is the charge of the electron, $\vec \Pi= \vec p -e \vec A$ and $\vec \sigma$ are the Pauli matrices.

\section{Finite Potential Barrier} \label{sec:fpb}

\begin{figure}
\centering
\includegraphics[scale=.5]{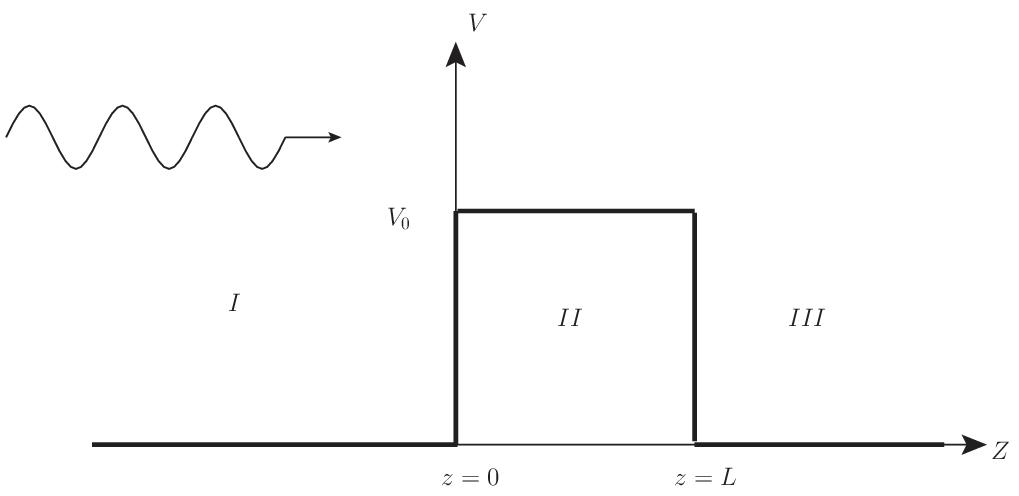}
\caption{Particle incident on a finite potential barrier of height $V_0$ and width $L$.}
\label{fig:barrier}
\end{figure}

In this section we analyze the finite potential barrier problem in 1 dimension (Figure \ref{fig:barrier}). The 1D equation is given by \cite{Ajaib:2015uha}
\begin{eqnarray}
-i \partial_z \psi = (i  \eta \partial_t  + \eta^\dagger m) \psi
\label{mse-eq-1d}
\end{eqnarray}
For the analysis in this section we adopt the representation of $\eta$ matrices employed in reference \cite{Ajaib:2015uha}.
We consider a spin up electron incident on a potential barrier with $E>V_0$ and $E<V_0$. We will show that the resulting transmission and reflection coefficients for spin up and down particles are related to the quantum mechanical ones. For regions $I$, $II$ and $III$, equation (\ref{mse-eq-1d}) in momentum space is given by
\begin{align}
p_1 \ \psi =& (E \eta  +m \eta^\dagger ) \ \psi  & (I, III) \label{eq:inc-wf} \\
p_2 \ \psi =& ((E-V_0) \eta  +m \eta^\dagger ) \ \psi & (II)  \label{eq:ref-wf}
\end{align}

\subsection{Case I: $E>V_0$}

\begin{figure}
%\vspace*{-1.2cm}
\centering
\includegraphics[scale=.5]{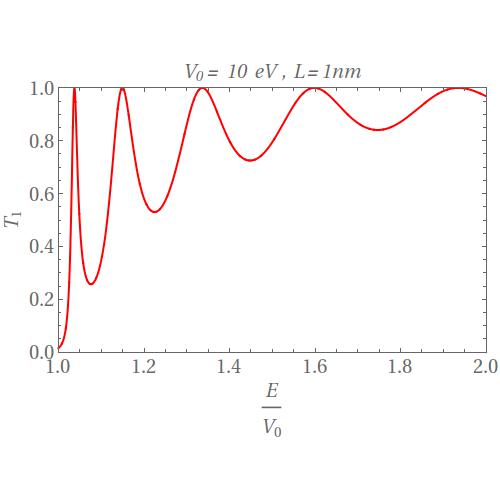}
\includegraphics[scale=.5]{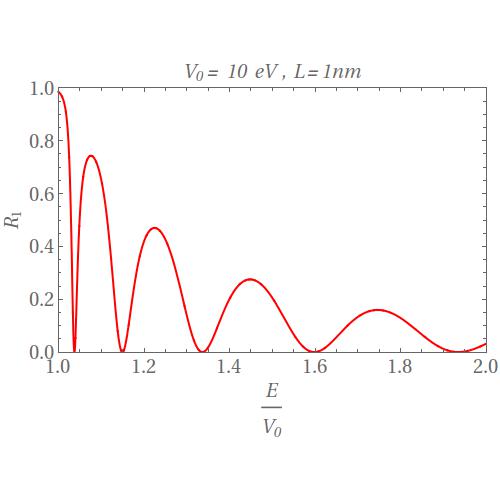}
\caption{The plot shows the transmission and reflection coefficients for spin up electrons given in equations (\ref{eq:t1}) and (\ref{eq:r1}). The height of the potential barrier $V_0$ is taken to be 10 eV and the width is chosen as $L=10$ nm. The transmission coefficient for spin down electron is zero whereas the reflection coefficient is negligible. For example, for $E/V_0=1.5$, the value of the reflection coefficient for spin down electron is $R_2=2.4\times10^{-5}$. The probability of the reflected electron to flip its spin however increases with increase in the energy of the particle. For instance, with $V_0=100$ KeV, $R_2=0.17$ for $E/V_0=1.5$. The analysis therefore predicts that for low energies the reflected electron will very likely be spin up  whereas for higher energies there is probability for it to be reflected with its spin flipped as well.
} 
\label{fig:transmission-10ev}
\end{figure}

We first consider the case of a spin up electron incident on a finite potential barrier of width $L$ with energy $E>V_0$. The incident, reflected and transmitted electron waves in region $I$, $II$ and $III$ are given by
\begin{eqnarray}
\psi_I=A \left(
\begin{array}{c}
 1 \\
 0 \\
 i \alpha (E-m) \\
 - \sqrt{2} \alpha  p_1 
\end{array}
\right) e^{i p_1 z}
+
A'\left(
\begin{array}{c}
 1 \\
 0 \\
 i \alpha (E-m) \\
  \sqrt{2} \alpha  p_1 
\end{array}
\right)  e^{-i p_1 z} +
B' \left(
\begin{array}{c}
 0 \\
 1 \\
  -  \sqrt{2}\alpha p_1 \\
 -i \alpha (E-m)
\end{array}
\right)  e^{-i p_1 z}
\end{eqnarray}

\begin{eqnarray}
\psi_{II}&=&F \left(
\begin{array}{c}
 1 \\
 0 \\
 i \beta (E-V_0-m) \\
 - \sqrt{2} \beta  p_2
\end{array}
\right) e^{i p_2 z}
+
F'\left(
\begin{array}{c}
 1 \\
 0 \\
 i \beta (E-V_0-m) \\
  \sqrt{2} \beta  p_2 
\end{array}
\right)  e^{-i p_2 z} \\
&+&G \left(
\begin{array}{c}
 0 \\
 1 \\
    \sqrt{2}\beta p_2 \\
 -i \beta (E-V_0-m)
\end{array}
\right)  e^{i p_2 z}
+
G' \left(
\begin{array}{c}
 0 \\
 1 \\
  -  \sqrt{2}\beta p_2 \\
 -i \beta (E-V_0-m)
\end{array}
\right)  e^{-i p_2 z}
\end{eqnarray}

\begin{eqnarray}
\psi_{III}=C\left(
\begin{array}{c}
 1 \\
 0 \\
 i \alpha (E-m) \\
  -\sqrt{2} \alpha  p_1 
\end{array}
\right)  e^{i p_1 z} +
D \left(
\begin{array}{c}
 0 \\
 1 \\
   \sqrt{2}\alpha p_1 \\
 -i \alpha (E-m)
\end{array}
\right)  e^{i p_1 z}
\end{eqnarray}

At $z=0$ and $z=L$ the continuity of the wave function implies
\begin{eqnarray}
\psi_I(z=0)&=&\psi_{II}(z=0) \\
\psi_{II}(z=L)&=&\psi_{III}(z=L)
\label{eq:cont-cond}
\end{eqnarray}
The transmission and reflection coefficients for spin up and down electrons for this case are
\begin{eqnarray}
T_1 &=& \left| \frac{C}{A} \right|^2 = \frac{8 E (E-V_0)}{8 E^2-V_0^2 \cos \left(2 \sqrt{2} L \sqrt{m (E-V_0)}\right)-8 E V_0+V_0^2} \label{eq:t1}
\\
T_2 &=& \left| \frac{D}{A} \right|^2 =0
\\
R_1 &=& \left| \frac{A'}{A} \right|^2 =\frac{2 V_0^2 (E-m)^2 \sin ^2\left(\sqrt{2} L \sqrt{m (E-V_0)}\right)}{(E+m)^2 \left(8 E^2-V_0^2 \cos \left(2 \sqrt{2} L \sqrt{m (E-V_0)}\right)-8 E V_0+V_0^2\right)} \label{eq:r1}
\\
R_2 &=& \left| \frac{B'}{A} \right|^2 =\frac{8 E m V_0^2 \sin ^2\left(\sqrt{2} L \sqrt{m (E-V_0)}\right)}{(E+m)^2 \left(8 E^2-V_0^2 \cos \left(2 \sqrt{2} L \sqrt{m (E-V_0)}\right)-8 E V_0+V_0^2\right)}
\end{eqnarray}
where $T_1$ and $R_1$ are transmission and reflection coefficients for spin up electron where as $T_2$ and $R_2$ correspond to spin down electron. The sum of these coefficient is always equal to 1, i.e.,
\begin{eqnarray}
 (T_1+T_2)+ (R_1+R_2) =1
\end{eqnarray}
The quantum mechanical transmission and reflection coefficients are related to these coefficients as 
\begin{eqnarray}
T_{QM} &=& T_1+T_2  \\
R_{QM} &=& R_1+R_2
\end{eqnarray}
Figure \ref{fig:transmission-10ev} shows the plot of transmission and reflection coefficients for spin up electrons in equations (\ref{eq:t1}) and (\ref{eq:r1}).  For these plots we choose $V_0=$10 eV and $L=10$ nm (with $m_ec^2=0.5$ MeV and $\hbar c=197$ eV-nm). From Figure \ref{fig:transmission-10ev} we can see that the transmitted electron is always spin up ($T_2=0$) whereas  the reflected electron is very likely spin up as well ($R_2<<1$).
For example, for $E/V_0=1.5$, the value of the coefficient for spin down electron is $R_2=2.4\times10^{-5}$. The probability of the reflected electron to flip its spin however increases with increase in the energy of the particle. For instance, with $V_0=100$ KeV, $R_2=0.17$ and $R_1=0.07$, for $E/V_0=1.5$. The analysis therefore predicts that for low energies the reflected electron will very likely be spin up  whereas for higher energies there is probability for it to be reflected with its spin flipped as well. The transmitted electron is always spin up.

\begin{figure}
%\vspace*{-1.2cm}
\centering
\includegraphics[scale=.5]{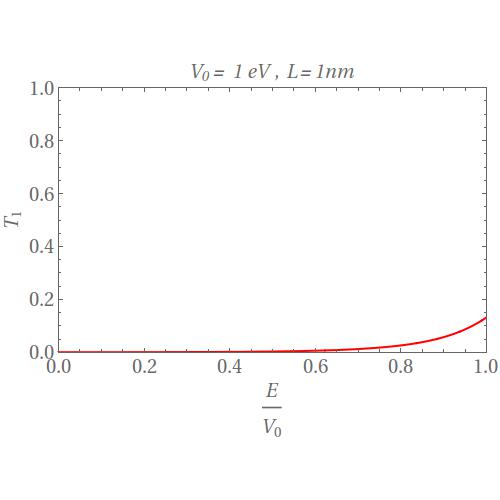}
\includegraphics[scale=.5]{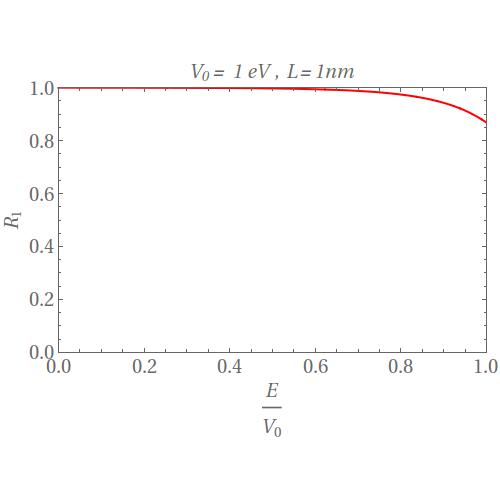}
\caption{The plot shows the transmission and reflection coefficients ($E<V_0$) for spin up electrons given in equations (\ref{eq:t1b}) and (\ref{eq:r1b}). We can see that there is a finite probability for the particle to tunnel through the barrier for higher energies. The transmitted electron is always spin up as well.}
\label{fig:reflection-1ev}
\end{figure}

\begin{figure}
%\vspace*{-1.2cm}
\centering
\includegraphics[scale=.35]{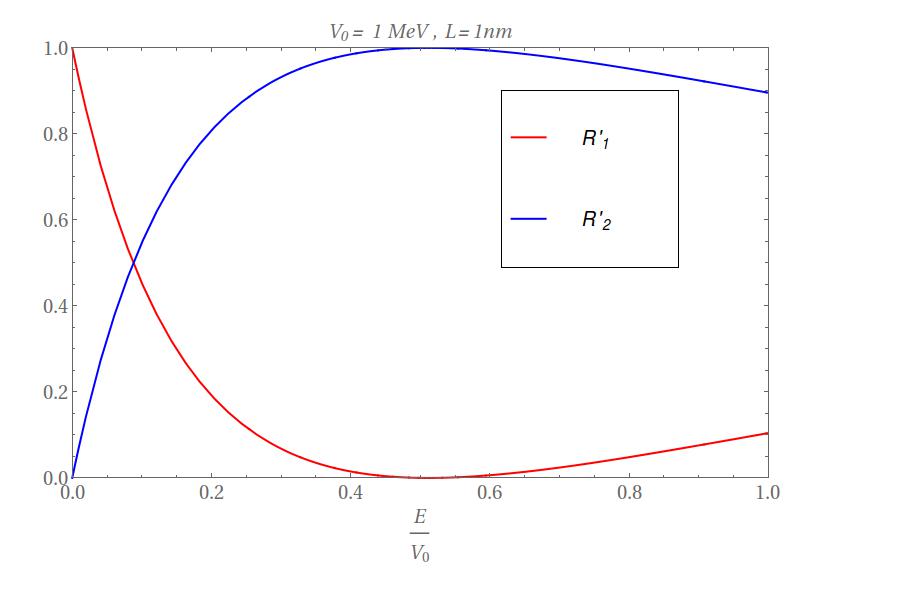}
\caption{The plot shows reflection coefficients for spin up (red) and spin down (blue) electrons given in equations (\ref{eq:r1b}) and (\ref{eq:r2b}). For these plots $V_0=$ 1 MeV. We can see that the probability of the electron to flip its spin upon reflection increases considerably for higher energies.}
\label{fig:reflection-1mev}
\end{figure}

\subsection{Case II: $E<V_0$}

We next analyze the case when the energy of the incident electron is less than the height of the barrier. The equations for incident and reflected electron in region $I$  remain the same as equations (\ref{eq:inc-wf}) and (\ref{eq:ref-wf}). 
In region $II$, however, the equation is now given by
\begin{align}
i p'_2 \ \psi =& [-(V_0-E) i\eta  +m i\eta^\dagger ] \ \psi & (II)  
\end{align}
The operator $\hat{P}=-(V_0-E) i\eta  +m i\eta^\dagger$ has eigenvalues $\pm i\sqrt{2(V_0-E)m}=\pm i p'_2$. The wave function in region $II$ is given by

\begin{eqnarray}
\psi_{II}&=&F \left(
\begin{array}{c}
 1 \\
 0 \\
 i  \rho (V_0-E+m) \\
 \sqrt{2} i \rho \  p'_2 
\end{array}
\right) e^{- p'_2 z}
+
F'\left(
\begin{array}{c}
 1 \\
 0 \\
 i  \rho (V_0-E+m) \\
   -\sqrt{2} i \rho \  p'_2  
\end{array}
\right)  e^{ p'_2 z} \\
&+&G \left(
\begin{array}{c}
 0 \\
 1 \\
    -\sqrt{2} i \rho \  p'_2  \\
 -i  \rho (V_0-E+m)
\end{array}
\right)  e^{- p'_2 z}
+
G' \left(
\begin{array}{c}
 0 \\
 1 \\
  \sqrt{2} i \rho \  p'_2  \\
 -i  \rho (V_0-E+m)
\end{array}
\right)  e^{ p'_2 z}
\end{eqnarray}
where $\rho=1/(V_0-E-m)$ and $p'_2=\sqrt{2m(V_0-E)}$. From the continuity condition (\ref{eq:cont-cond}) we find the coefficients for this case as well:

\begin{eqnarray}
T'_1 &=& \left| \frac{C}{A} \right|^2 = \frac{8 E (E-V_0)}{8 E^2-V_0^2 \cosh \left(2 \sqrt{2} L \sqrt{m (V_0-E)}\right)-8 E V_0+V_0^2} \label{eq:t1b}
\\
T'_2 &=& \left| \frac{D}{A} \right|^2 =0
\\
R'_1 &=& \left| \frac{A'}{A} \right|^2 =-\frac{2 V_0^2 (E-m)^2 \sinh ^2\left(\sqrt{2} L \sqrt{m (V_0-E)}\right)}{(E+m)^2 \left(8 E^2-V_0^2 \cosh \left(2 \sqrt{2} L \sqrt{m (V_0-E)}\right)-8 E V_0+V_0^2\right)}  \label{eq:r1b}
\\
R'_2 &=& \left| \frac{B'}{A} \right|^2 =-\frac{8 E m V_0^2 \sinh ^2\left(\sqrt{2} L \sqrt{m (V_0-E)}\right)}{(E+m)^2 \left(8 E^2-V_0^2 \cosh \left(2 \sqrt{2} L \sqrt{m (V_0-E)}\right)-8 E V_0+V_0^2\right)} \label{eq:r2b}
\end{eqnarray}

The sum of these coefficient also is always equal to 1, i.e.,
\begin{eqnarray}
 (T'_1+T'_2)+ (R'_1+R'_2) =1
\end{eqnarray}
with the coefficients from the Schr{\"o}dinger equation given by $T_{QM} = T'_1+T'_2$ and $R_{QM} = R'_1+R'_2$.

In Figures \ref{fig:reflection-1ev} and \ref{fig:reflection-1mev} we show the results for this case. In Figure \ref{fig:reflection-1ev} we choose $V_0=$1 eV and $L=10$ nm and in Figure \ref{fig:reflection-1mev}, $V_0=$1 MeV. From Figure \ref{fig:reflection-1ev} we can see that there is a finite probability for the particle to tunnel across the barrier, especially for higher energies.  For larger heights of the potential barrier the transmission coefficient decreases considerably. The transmitted electron is always spin up whereas the reflected electron can be spin up or down depending on the energy of the particle. The probability of the electron to flip its spin upon reflection increases considerably for higher energies. This can be seen from Figure \ref{fig:reflection-1mev} which shows the reflection coefficients of the spin up (red) and down (blue) electron.

\section{Particle in a Symmetric Well} \label{sec:box}

In this section we address the question of confinement of a particle in a 1 dimensional symmetric potential well ($z:-L \rightarrow L$) using equation (\ref{mse-eq-1d}). We find that in order to confine the particle in a symmetric potential well we need to implement periodic boundary conditions.
In the potential well the particle is described by the following wave function:
\begin{eqnarray}
\psi=A \left(
\begin{array}{c}
 1 \\
 0 \\
 i \alpha (E-m) \\
 - \sqrt{2} \alpha  p_1 
\end{array}
\right) e^{i p_1 z}
+
B \left(
\begin{array}{c}
 0 \\
 1 \\
    \sqrt{2}\alpha p_1 \\
 -i \alpha (E-m)
\end{array}
\right)  e^{i p_1 z}
\nonumber \\ 
+
A'\left(
\begin{array}{c}
 1 \\
 0 \\
 i \alpha (E-m) \\
  \sqrt{2} \alpha  p_1 
\end{array}
\right)  e^{-i p_1 z} +
B' \left(
\begin{array}{c}
 0 \\
 1 \\
  -  \sqrt{2}\alpha p_1 \\
 -i \alpha (E-m)
\end{array}
\right)  e^{-i p_1 z}
\end{eqnarray}
The wave function $\psi$ is a linear combination of eigenstates of spin up and down particles moving towards the positive and negative $z$ axis. For this case we implement periodic boundary conditions
\begin{eqnarray}
\psi(z=-L)&=&\psi(z=L) 
\label{eq:cont-cond-symm}
\end{eqnarray}
The allowed quantized energy levels of the particle are given by
\begin{eqnarray}
E_n=\frac{n^2 \pi^2 }{2 m L^2}
\end{eqnarray}
where $n=1,2,3, ...$. These are the energy levels of particle in a square well.
Note that in this case the coefficients ($A$, $B$, $A'$ $B'$) cannot be determined, however, the normalization of the wave function 
\begin{eqnarray}
\int^L_{-L}|\psi|^2 dz=1
\end{eqnarray}
yields the following condition on the coefficients of the wave function
\begin{eqnarray}
|A|^2+|B|^2+|A'|^2+|B'|^2=\frac{1}{4L}.
\end{eqnarray}

\section{Conclusion} \label{conclude}

We have shown that the equation proposed in \cite{Ajaib:2015uha} is the non-relativistic limit of the Dirac equation. We also showed that the Pauli equation can be derived by requiring this equation to be locally invariant. 
In addition, we analyzed the finite potential barrier problem and showed that in this case as well the transmission and reflection coefficients for spin up and down electrons, when added together, yield the quantum mechanical results for this problem. The analysis predicts that a spin up electron incident on a finite potential barrier is always transmitted as spin up. For low energies the reflected electron is most likely spin up as well,  whereas for higher energies there is a greater probability for it to be reflected with its spin flipped as well.

Finally, we considered a particle in a symmetric potential well and showed that quantized energy levels are obtained with periodic boundary conditions. The results obtained from the equation proposed in \cite{Ajaib:2015uha} agree well with known quantum mechanical expressions. The analyses performed in \cite{Ajaib:2015uha} and in this article predict precisely the reflection and transmission coefficients for spin up and down electrons for the step potential and finite potential barrier problems. Experimental tests are needed to examine the predictions of these analyses.

Based on these conclusions, we suggest that this equation more appropriately describes spin 1/2 particles in the non-relativistic limit.

\section{Acknowledgements}

The author would like to thank Fariha Nasir, Warren Siegel, Mansoor Ur Rehman and Robert Echols for useful discussions and suggestions. The author is also grateful to the referees of \cite{Ajaib:2015uha} for useful suggestions.


\begin{thebibliography}{99}

%\cite{Ajaib:2015uha}
\bibitem{Ajaib:2015uha} 
  M.~A.~Ajaib,
  %``A Fundamental Form of the Schr{\"o}dinger Equation,''
  Found.\ Phys.\  {\bf 45}, no. 12, 1586 (2015)
  doi:10.1007/s10701-015-9944-z
  [arXiv:1502.04274 [quant-ph]].
  %%CITATION = doi:10.1007/s10701-015-9944-z;%%



\end{thebibliography}
\end{document}